%% file: revised.tex
\documentclass[useAMS,usenatbib]{mn2e}

\usepackage{graphicx}
\usepackage{amsmath}
\usepackage[dvips,usenames]{color}
\input{journalDefs}

\usepackage{times}

\newcommand{\vi}{\mbox{$V\!-\!I$}}

\newcommand{\Msun}{\mbox{$M_{\odot}$}}
\newcommand{\Teff}{\mbox{$T_{\rm eff}$}}

\newcommand{\comment}[1]{}
\newcommand{\beq}{\begin{equation}}
\newcommand{\eeq}{\end{equation}}
\newcommand{\beqa}{\begin{eqnarray}}
\newcommand{\eeqa}{\end{eqnarray}}

\title[Rotation and MMSTOs]
{Can rotation explain the multiple main sequence turn-offs of
Magellanic Cloud star clusters?}

\author[Girardi et al.]{
	L\'eo Girardi$^{1}$,
	Patrick Eggenberger$^{2}$,
	Andrea Miglio$^{3}$ \\
$^{1}$ Osservatorio Astronomico di Padova -- INAF,
	Vicolo dell'Osservatorio 5, I-35122 Padova, Italy \\
$^{2}$ Observatoire de Gen\`eve, Universit\'e de Gen\`eve, 
	51 chemin des Maillettes, 1290 Sauverny, Switzerland \\
$^{3}$ Institut d'Astrophysique et de G\'eophysique de 
	l'Universit\'e de Li\`ege, All\'ee du 6 Ao\^ut, 17 B-4000 
	Li\'ege, Belgium 
}

\begin{document}

\date{To appear in MNRAS Letters}

\pagerange{\pageref{firstpage}--\pageref{lastpage}} \pubyear{2011}

\maketitle

\label{firstpage}

\begin{abstract}
Many intermediate age star clusters in the Magellanic Clouds present
multiple main sequence turn-offs (MMSTO), which challenge the
classical idea that star formation in such objects took place over
short timescales. It has been recently suggested that the presence of
fast rotators among main sequence stars could be the cause of such
features \citep{BastiandeMink09}, hence relaxing the need for extended
periods of star formation. In this letter, we compute evolutionary
tracks and isochrones of models with and without rotation. We find
that, for the same age and input physics, both kinds of models present
turn-offs with an almost identical position in the colour-magnitude
diagrams. As a consequence, a dispersion of rotational velocities in
coeval ensembles of stars could not explain the presence of MMSTOs. We
construct several synthetic colour-magnitude diagrams for the
different kinds of tracks and combinations of them. The models that
best reproduce the morphology of observed MMSTOs are clearly those
assuming a significant spread in the stellar ages -- as long as
$\sim400$~Myr -- added to a moderate amount of convective core
overshooting. Only these models produce the detailed ``golf club''
shape of observed MMSTOs. A spread in rotational velocities alone
cannot do anything similar. We also discuss models involving a mixture
of stars with and without overshooting, as an additional scenario to
producing MMSTOs with coeval populations. We find that they produce
turn-offs with a varying extension in the CMD direction perpendicular
to the lower main sequence, which are clearly not present in observed
MMSTOs.
\end{abstract}


\section{Introduction}
\label{intro}

There is now conclusive evidence that many star clusters in the
Magellanic Clouds present double or multiple main sequence turn-offs
(MMSTO). The first hints of this phenomenon have been advanced by
\citet{Bertelli_etal03} and \citet{Baume_etal07}, and were based on
ground-based data for the LMC clusters NGC~2173 and NGC~2154. Much
more impressive and conclusive, however, were the evidences brought by
\citet{Mackey_BrobyNielsen2007}, \citet{Mackey_etal08},
\citet{Goudfrooij_etal09} for the clusters
NGC~1846, NGC~1806, and NGC~1783 in the LMC, and \citet{Glatt_etal08}
for NGC~419 in the SMC, based on much deeper photometry obtained with
the Advanced Camera for Surveys (ACS) onboard the Hubble Space
Telescope (HST). In all these cases, the presence of a broad or MMSTO
clearly stands out even from a visual inspection of the
CMDs. \citet{Milone_etal08} revised the available ACS photometry for
16 intermediate-age LMC clusters, finding evidence for broad turn-offs
in about 70~\% of them. 

The easiest interpretation of MMSTOs is that these clusters contain
two or more generations of stars formed one after the other, over a
time span of a few hundreds of Myr. This is indeed the interpretation
adopted by the above-mentioned works. In support of this
interpretation, there is the fact that a subsample of the
intermediate-age clusters contain a dual clump of red giants
\citep{Girardi_etal09}, indicative of both the onset of electron
degeneracy in stellar cores, and of a $\sim0.15$~\Msun\ spread in
turn-off masses which is compatible with the one indicated by the
color spread in the turn-off region \citep[][and work in
preparation]{Rubele_etal10}.

This interpretation however is not an easy one, when one considers the
challenges it poses to the theory of star and cluster formation
\citep[see e.g.][]{Goudfrooij_etal09, ConroySpergel10}. Note also that
the prolonged star formation probably occurred {\em in situ}, within
the relatively shallow potential well of $\la10^5$~\Msun\ clusters,
and are not due to the merging of different clusters
\citep[][]{Goudfrooij_etal09}.

\begin{figure*}
\resizebox{0.33\hsize}{!}{\includegraphics{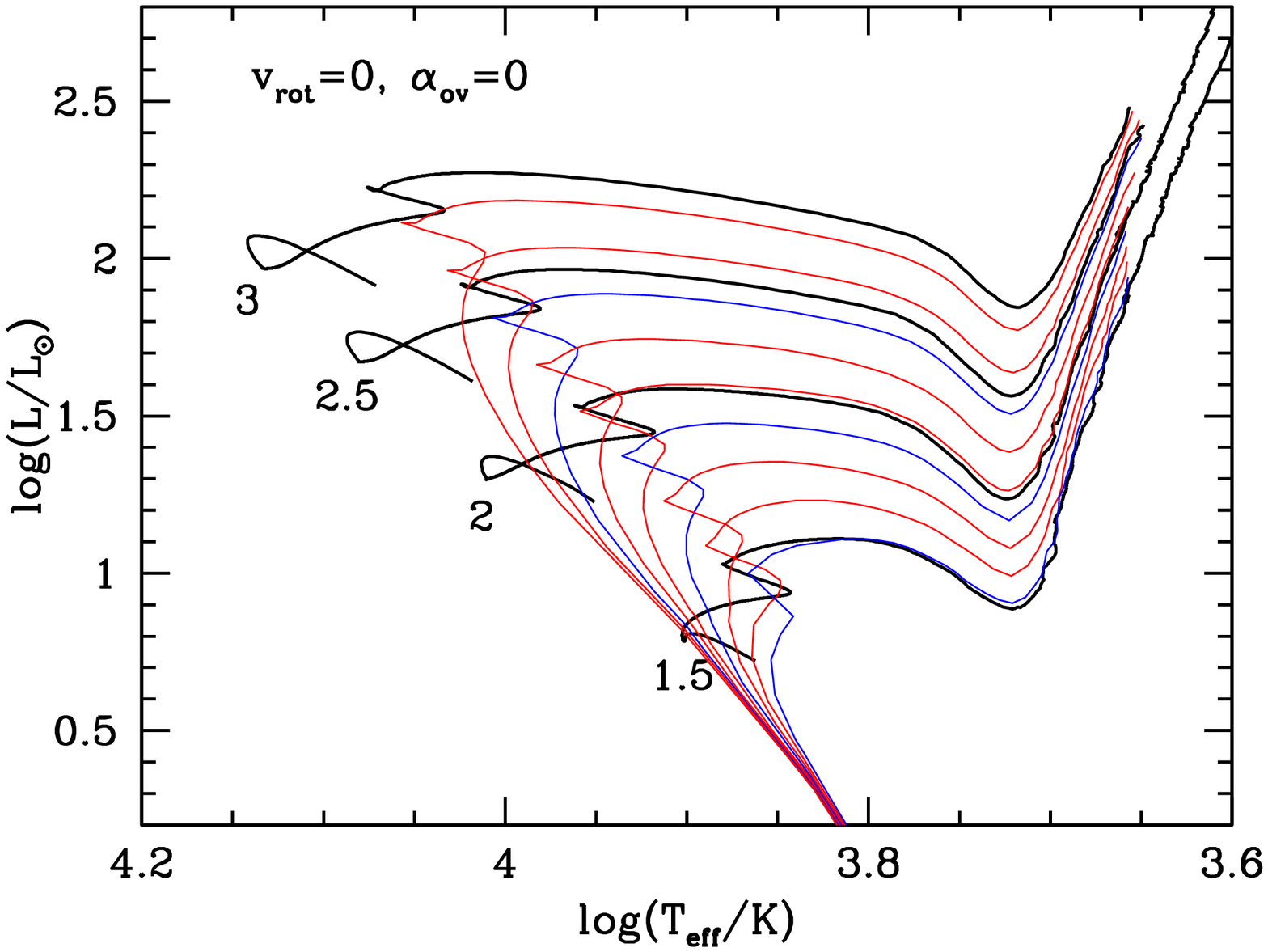}}
\resizebox{0.33\hsize}{!}{\includegraphics{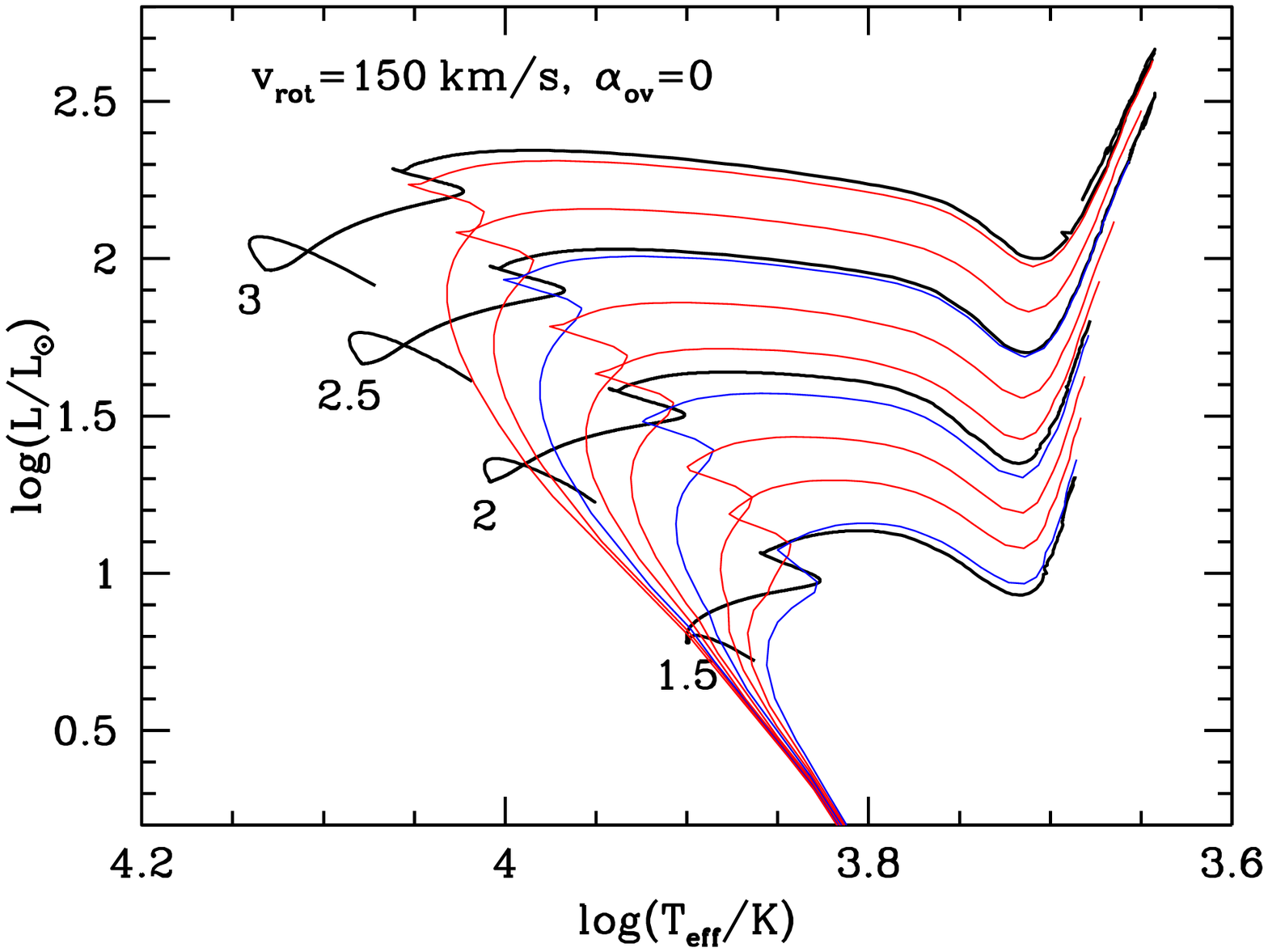}}
\resizebox{0.33\hsize}{!}{\includegraphics{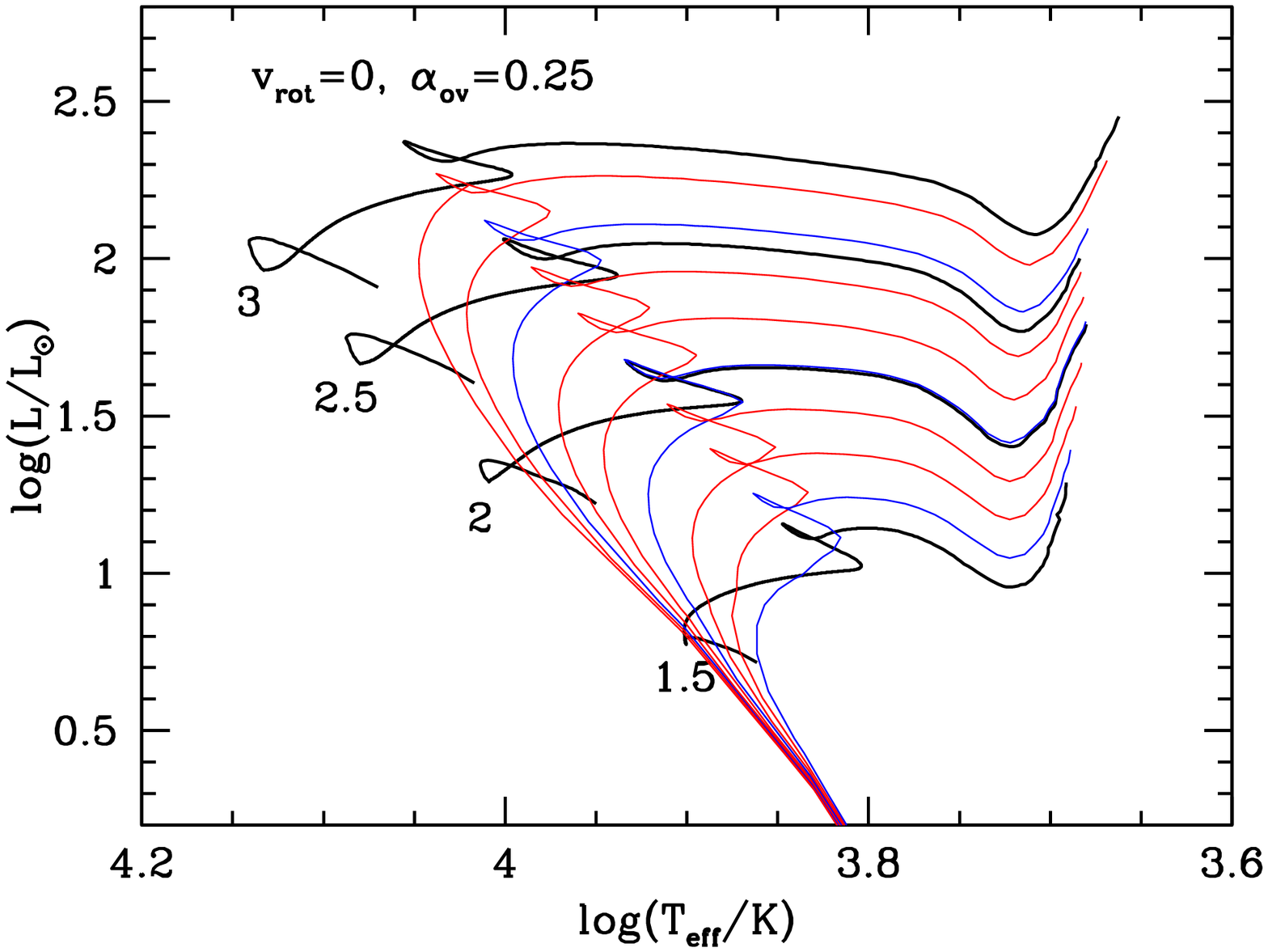}}
\caption{The evolutionary tracks computed for this work (dark solid lines), 
together with isochrones derived from them (narrow lines, in color in
the electronic version), for the non-rotating, rotating, and
overshooting cases (from left to right, respectively). The isochrones
ages go, from top to bottom, from $\log(t/{\rm yr})=8.5$ to 9.3 at
steps of 0.1~dex. As a reference to the eye, the isochrones with ages
of 0.5, 1 and 2~Gyr are marked with a different colour.}
\label{fig_tracks}
\end{figure*}

There is, however, an amazing coincidence regarding the age scales in
all these cases. For all LMC clusters with the clear presence of
MMSTOs in the above-mentioned papers, the age interval between the
youngest and oldest turn-off ranges from about 150 to $\sim300$~Myr,
whereas the clusters themselves have all ages between 1 and 2~Gyr
\citep[see][]{Milone_etal08}. NGC~419 in the SMC indicates a maximum
age spread of 700~Myr, but again for a mean age of 1.5~Gyr
\citep{Rubele_etal10}. In terms of CMD features, the MMSTOs 
always cover a magnitude range of about $\sim0.5$~mag in passbands
like $V$ or $I$, and $\sim0.15$~mag in colours like $\vi$. The MMSTO
phenomenum is observed in star clusters located overall over the LMC
disk and also in the SMC, although it seems to be preferentially
observed in the most massive objects \citep{ConroySpergel10}. The
homogeneity in the mean ages of clusters with MMSTOs could be
indicating dramatic large-scale events in the past history of the
Magellanic Clouds, that could have affected the formation of many of
its more massive clusters in a similar way.
Or, perhaps, this coincidence is simply suggesting us to look for a
completely different interpretation to the MMSTO phenomenum.

\citet{BastiandeMink09} have so far advanced the only alternative
explanation for such CMD features, namely that the MMSTOs appear in
aggregates of {\em coeval} stars, due to the color dispersion caused
by the presence of fast rotators among the main sequence stars. Their
suggestion is based on simple considerations about the shape of
stellar evolutionary tracks with and without rotation, on some
observational facts, and on some less well-justified assumptions,
namely that:
\begin{itemize}
\item a significant fraction of main sequence stars rotate with 
velocities of about $\sim0.4$ times the critical break up velocity,
$\Omega_{\rm crit}$ \citep[see][]{Royer_etal07};
\item rotation causes stars to shift, in general, to redder colours 
and slightly brighter luminosities in CMDs;
\item for the most massive stars, close to the MSTO, the effect of 
rotation is quite similar throughout the main-sequence evolution, so
that the effects are considered as displayed when stars reach the end
of the main sequence;
\item the effect of rotation gets less important at smaller masses, 
and disappears for masses smaller than $\sim1.2$~\Msun.
\end{itemize}
Note that {\em the combination} of all these different effects is
required to explain the shape of the observed MMSTOs. In addition,
strongly bi-modal distributions of rotational velocities are needed to
explain the few cases in which the turn-off appear as a double feature.

In this paper, we investigate whether rotation does really behave in
the way suggested by \citet{BastiandeMink09}. Our analysis is based on
isochrones computed with detailed models of rotating stars
(Sect.~\ref{sec_tracks}), and do not suffer from the approximations
adopted by them. In Sect.~\ref{sec_discussion} we compare the
performance of coeval models with a dispersion of rotation, with that
of non rotating models with a dispersion of ages, in producing the
MMSTO feature observed in the LMC cluster NGC~1846. The basic
conclusions are discussed in Sect.~\ref{sec_conclu}.

\section{Models of rotating stars and isochrones}
\label{sec_tracks}

We start computing a series of evolutionary tracks with initial masses
in the interval relevant for MMSTOs. An initial heavy elements mass
fraction of $Z=0.008$, suitable for LMC populations of
young-to-intermediate ages, is adopted.

The main-sequence evolution of models with and without rotation is
computed by using the Geneva stellar evolution code \citep{egg08}. The
standard mixing-length formalism for convection \citep{boe58} is used
with a solar calibrated value of the mixing-length
parameter. Overshooting from the convective core into the surrounding
radiatively stable layers is not included. Rotating models are
computed with an initial velocity on the ZAMS of 150~km\,s$^{-1}$,
which corresponds to a typical value for rotating stars in the mass
interval between 1.5 and 3~\Msun\ studied here
\citep[e.g.][]{Royer_etal07}. \footnote{Note however that for the lower
metallicities of the LMC and SMC, the median rotation velocity might
be somewhat higher than in the Solar Neighbourhood sample from
\citet{Royer_etal07}. This difference however is unlikely to affect
the main arguments discussed in the following.}
Note that the ratio $\omega=\Omega / \Omega_{\rm crit}$ of these
models remains always inferior to 0.7, so that the effects related to
the variation with colatitude of the effective temperature are
negligible \citep{Ekstrom_etal08}. These models include a
comprehensive treatment of shellular rotation
\citep{zah92} and meridional circulation is treated as a truly
advective process \citep[for more details, see e.g.][]{egg10}. The
rotating and non-rotating models share the same initial parameters
except for the inclusion of shellular rotation.

These two sets are complemented by one without rotation but with
overshooting from the convective core, computed with the same
code. The overshooting parameter is set to $\alpha_{\rm ov}=0.25$
pressure scale heights (see \citealt{MaederMeynet89} for details).

The tracks are presented in Fig.~\ref{fig_tracks}. They cover all
evolutionary stages from the zero-age main sequence (ZAMS) up to
either He-ignition, or the RGB-bump of less massive stars.

\begin{figure*}
\resizebox{0.33\hsize}{!}{\includegraphics{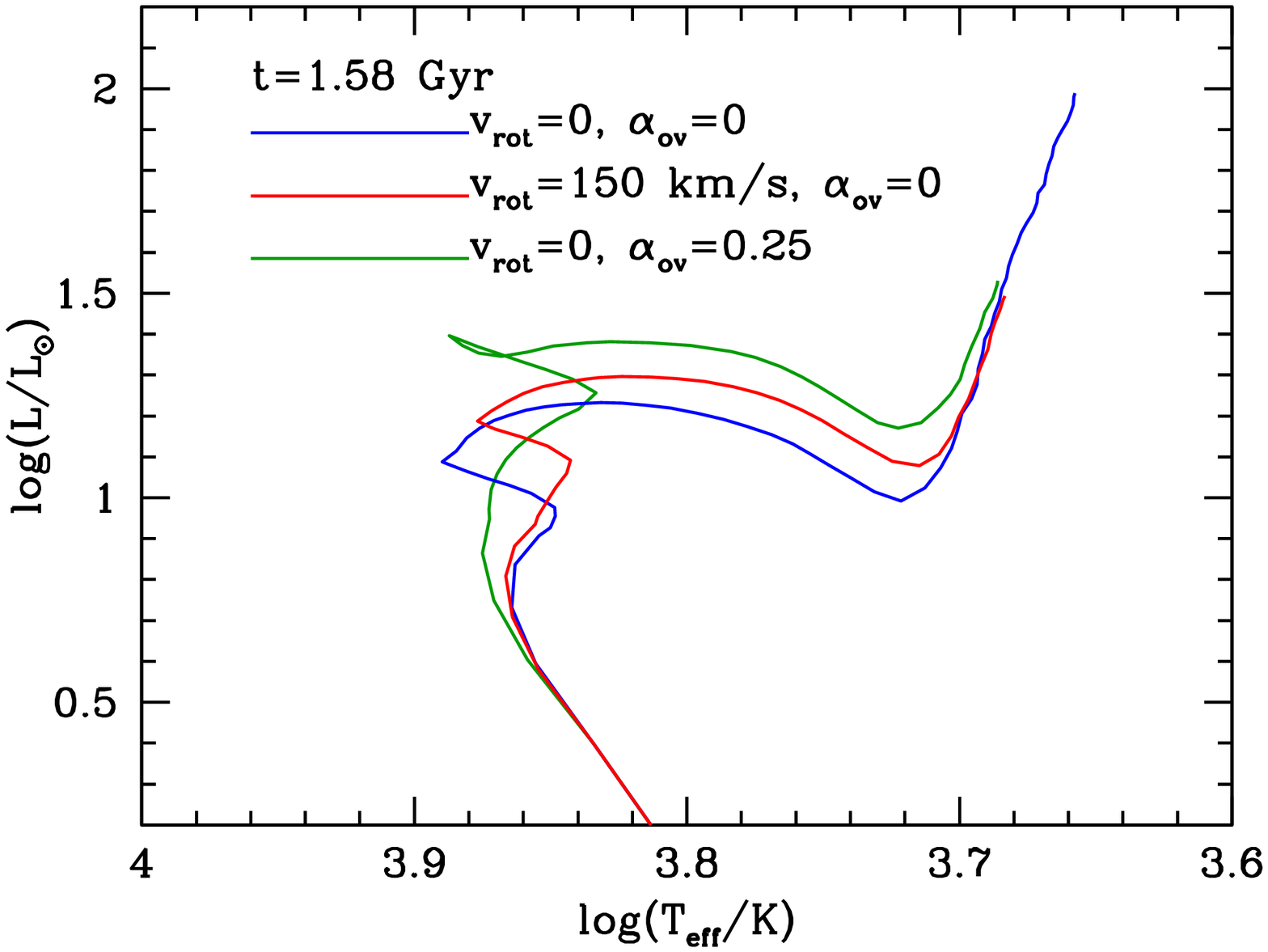}}
\resizebox{0.33\hsize}{!}{\includegraphics{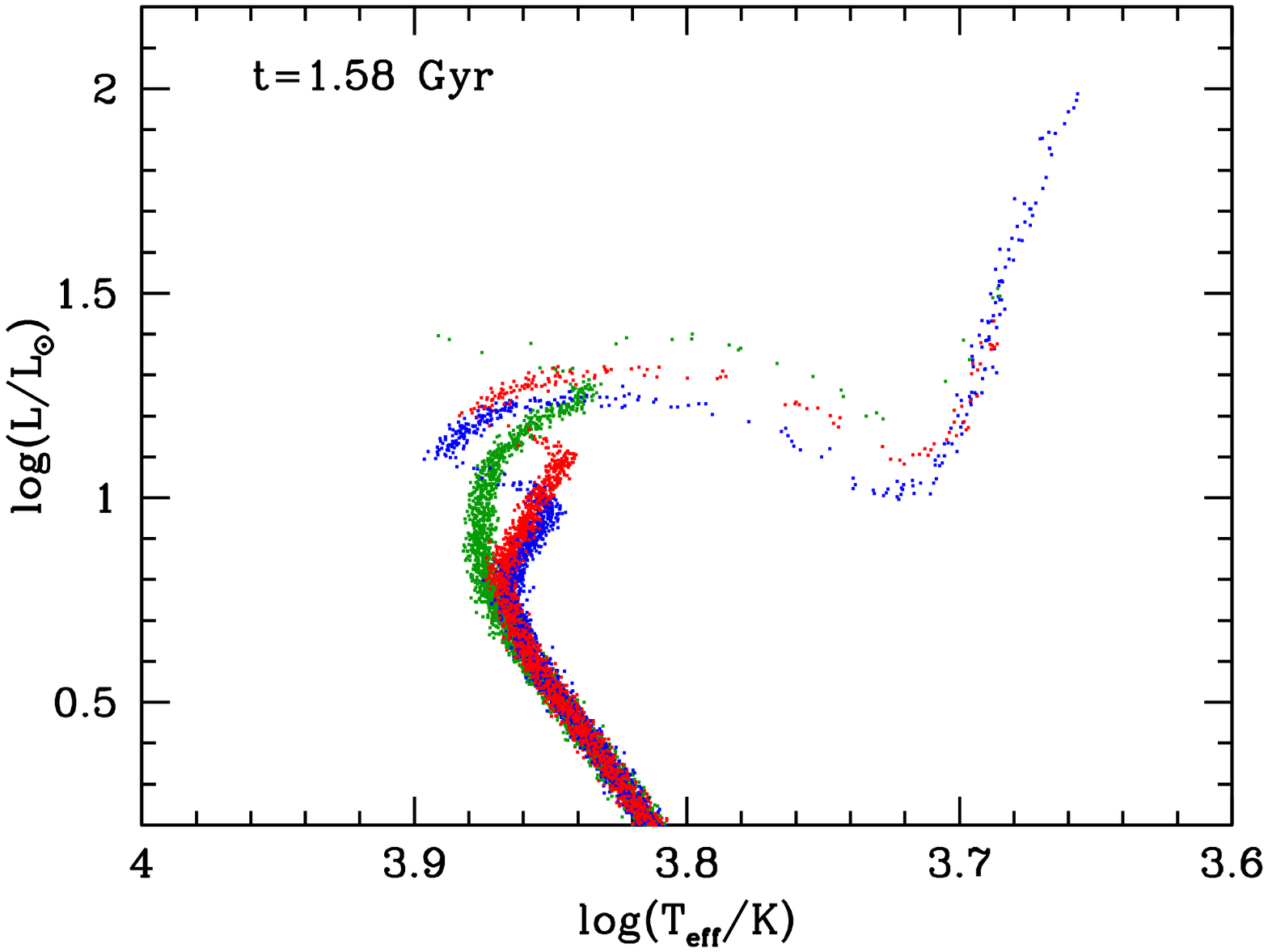}}
\caption{Left panel: HR diagram of the different isochrones obtained 
in Fig.~\ref{fig_tracks}, for the same age of 1.58~Gyr. Right: a
synthetic HR diagram where the same isochrones are populated by
$10^4$~\Msun\ following a \citet{Chabrier01} log-normal IMF. To avoid
the excessive superposition of points in the diagram, they were
displaced by random Gaussian errors of 0.01~dex in $\log L$ and
0.0025~dex in $\log T_{\rm eff}$.}
\label{fig_isocs}
\end{figure*}

The same figure presents a set of isochrones derived from
them. Despite the relatively large mass interval between successive
tracks, the interpolation algorithm \citep[the same as
in][]{Girardi_etal00} works very well and produces isochrones with all
the fine details of the original tracks, especially along the MSTO
region.  In order to include the lower main sequence in the
isochrones, the new grids of tracks have been complemented with
\citet{Girardi_etal00} non-rotating tracks of $M\le1.0$~\Msun, which, 
for the ages considered in this work, are always very close to the
ZAMS.

The two sets of tracks without overshooting make the basic set with
which we will discuss the effect of rotation on the CMDs. We recall
that these tracks do follow, qualitatively, the changes in the stellar
$\Teff$ and $L$ due to rotation that are at the base of
\citet{BastiandeMink09} work.

\section{Synthetic CMDs and comparison with data}
\label{sec_discussion}

Let us now consider the effect that rotation has in the CMDs of star
clusters. Let us assume for the moment that all the stars in clusters
with multiple turn-offs have exactly the same age and initial chemical
composition. The theoretical entities suitable to describe such a
situation are the stellar isochrones. In the following, we will center
our discussion on isochrones computed for an age of 1.58~Gyr, however
we advance that very similar conclusions would be obtained for any age
between 1 and 2 Gyr.

In the left panel of Fig.~\ref{fig_isocs}, we superpose isochrones of
the same age computed for the 3 cases mentioned in
Sect.~\ref{sec_tracks}, in the HR diagram. The result is somewhat
surprising: contrarily to what sustained in \citet{BastiandeMink09},
{\em models with rotation do not present a cooler and fainter
turn-off\footnote{Throughout this paper, turn-off is intended as the
tip of the locus {\em continuously} populated by stars along the main
sequence.} with respect to their non-rotating counterparts. Instead,
they have a slightly hotter and brighter turn-off}.  This result
cannot be easily inferred while looking only at the evolutionary
tracks in Fig.~\ref{fig_tracks}, which at first sight seem to suggest
that rotating models have a broader main sequence, hence being able to
extend to cooler temperatures, as sustained in
\citet{BastiandeMink09}. This is true only for pairs of evolutionary
tracks of the same mass. What we see in the isochrones is different,
it is the {\em combined effect} of changes in the morphology of the
evolutionary tracks, and of the changes in their lifetimes.

The increase in the main sequence lifetimes caused by rotation, in
particular, is quite remarkable \citep[see][]{Palacios_etal03, egg10}
and constitutes one of the main factors in play here. As discussed in
\citet{egg10}, classical models with a $v_{\rm rot}=150$~km/s have a 
similar increase in their main sequence lifetimes as models with a
moderate amount of overshooting, that is $\alpha_{\rm ov}=0.1$. Such
an increase in the lifetimes of rotating stars has been ignored in
\citet{BastiandeMink09}, and this has been determinant in leading to
their different conclusions.

In the right panel of Fig.~\ref{fig_isocs}, we present synthetic HR
diagrams corresponding to 1.58-Gyr old clusters with and without
rotation -- in this latter case assuming that all stars rotate with
the same initial velocity -- and with overshooting, for a total mass
of $10^4$~\Msun\ following the \citet{Chabrier01} log-normal IMF. This
panel clearly evinces what are the regions of the diagram effectively
populated by stars. In particular, one may notice that the differences
between non-rotating and rotating models is not dramatic, if one
considers only the region where the MSTO is drawn.

\begin{figure*}
\resizebox{0.33\hsize}{!}{\includegraphics{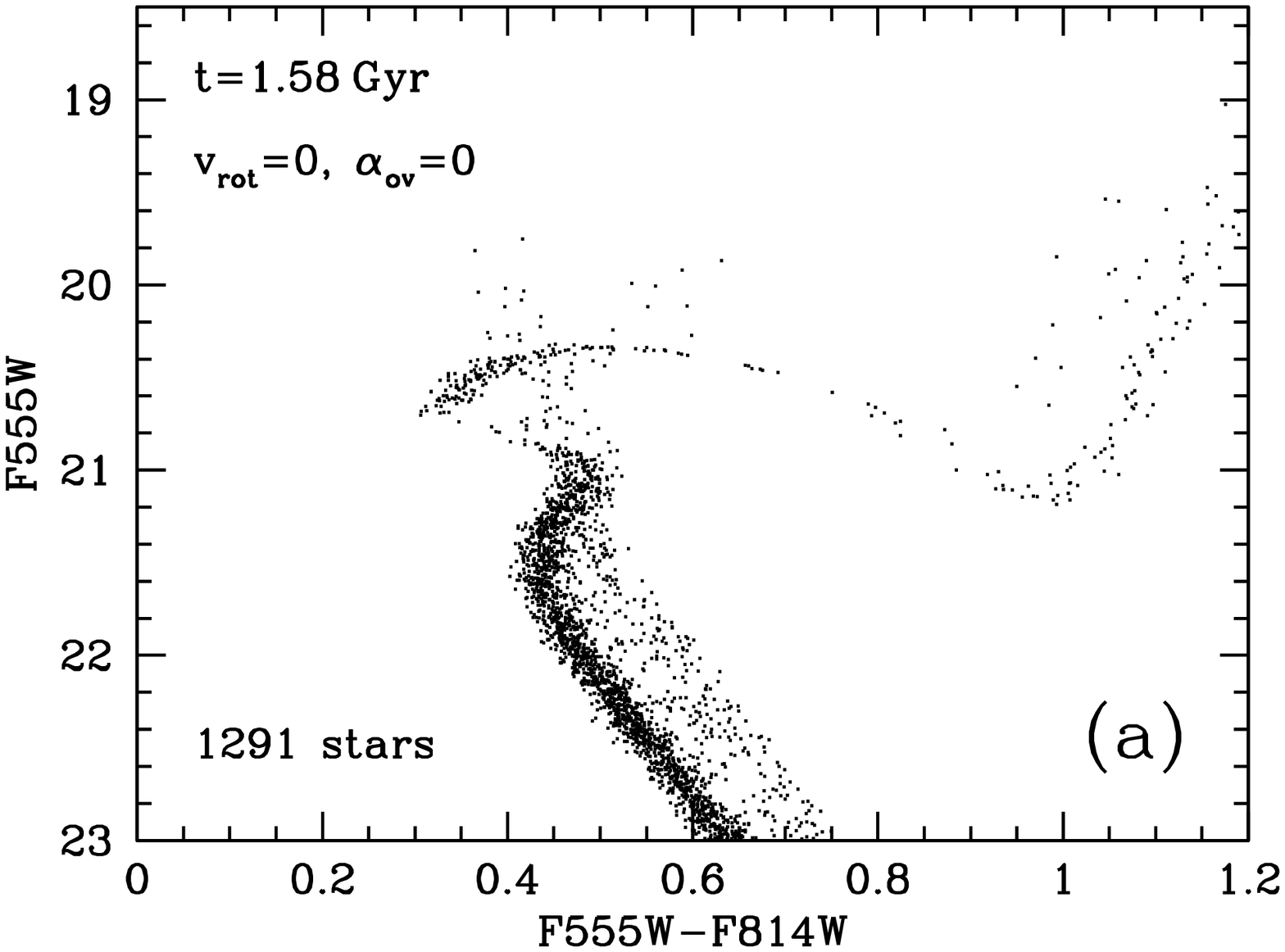}}
\resizebox{0.33\hsize}{!}{\includegraphics{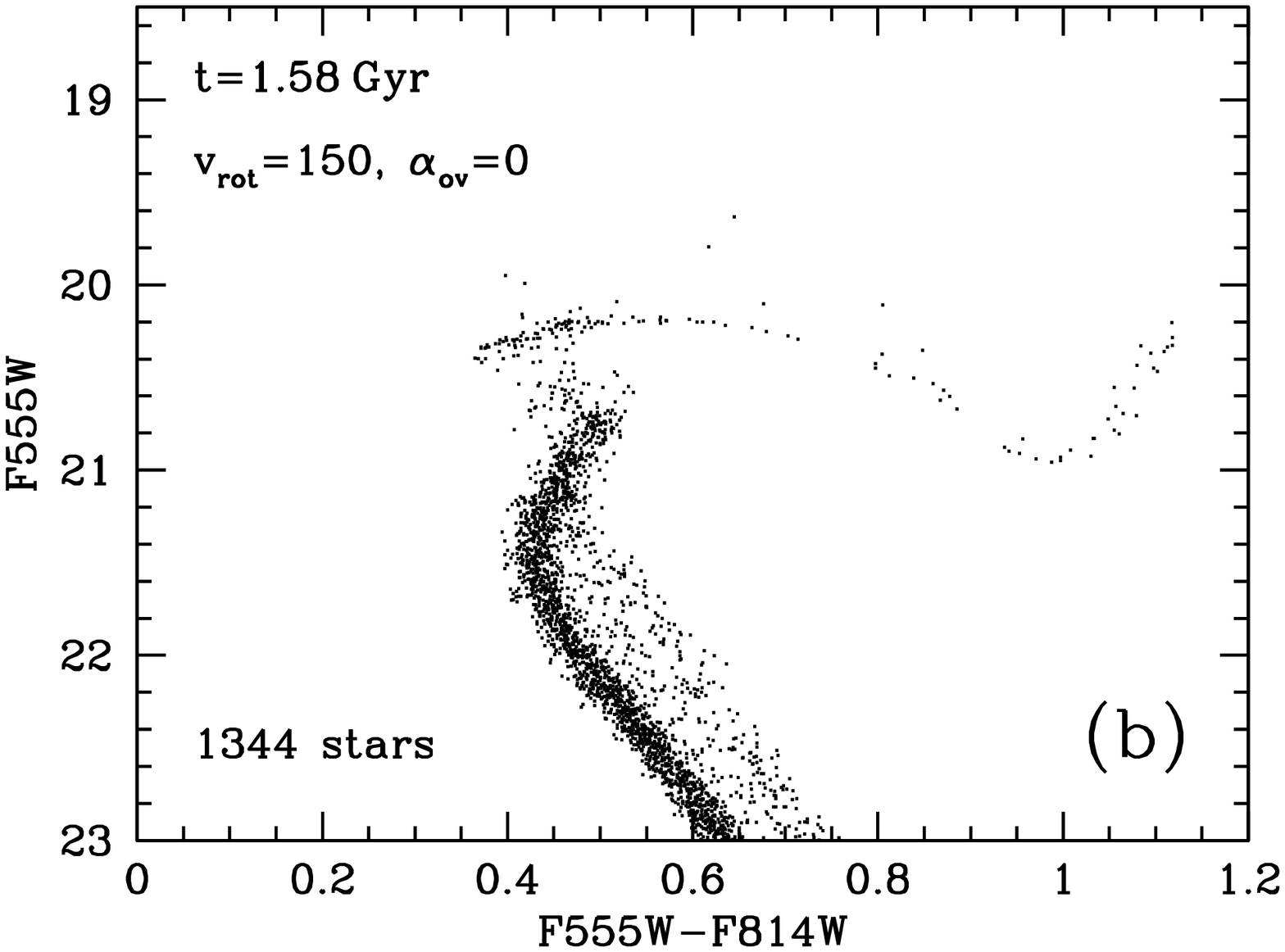}}
\resizebox{0.33\hsize}{!}{\includegraphics{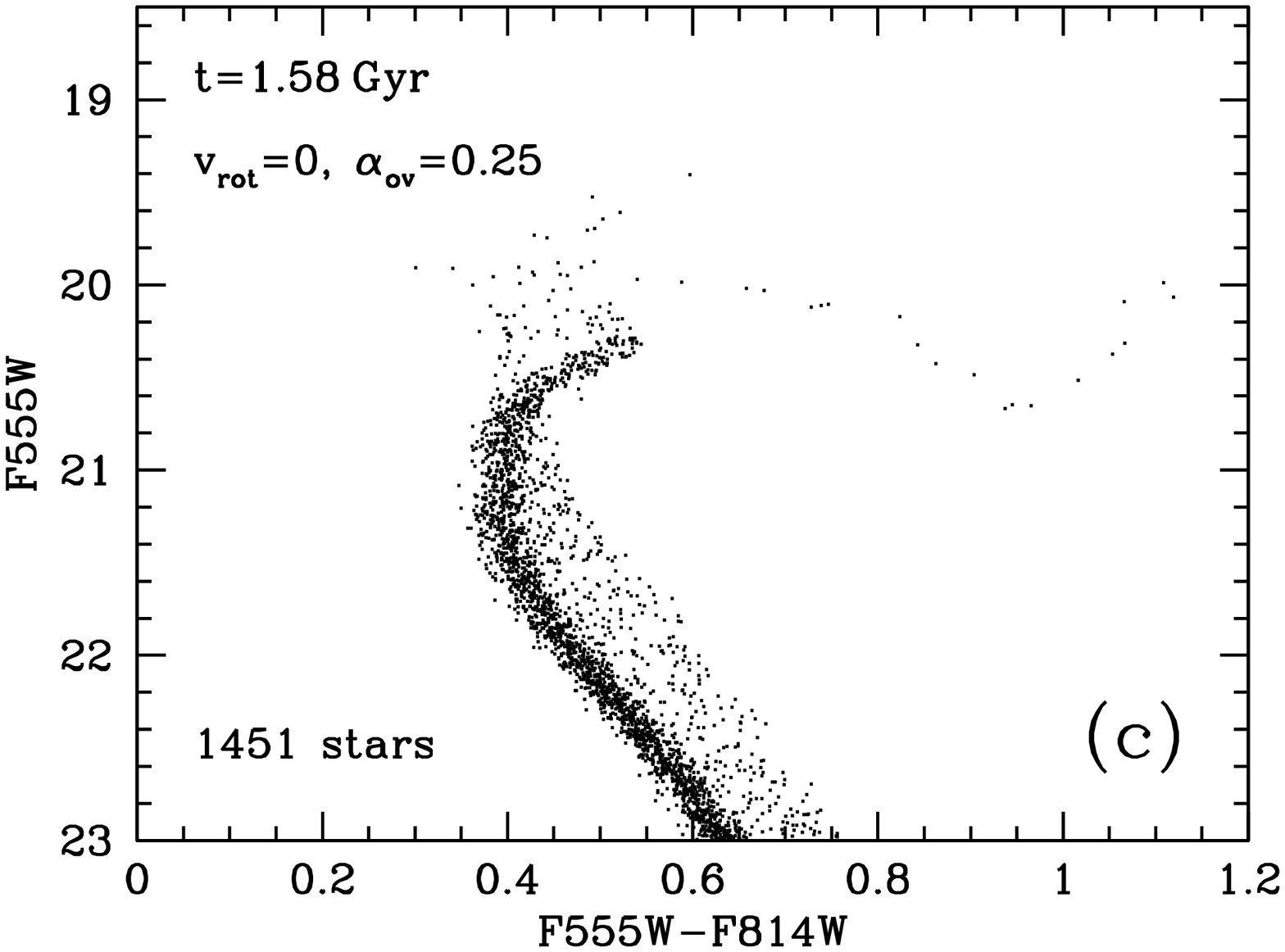}}
\resizebox{0.33\hsize}{!}{\includegraphics{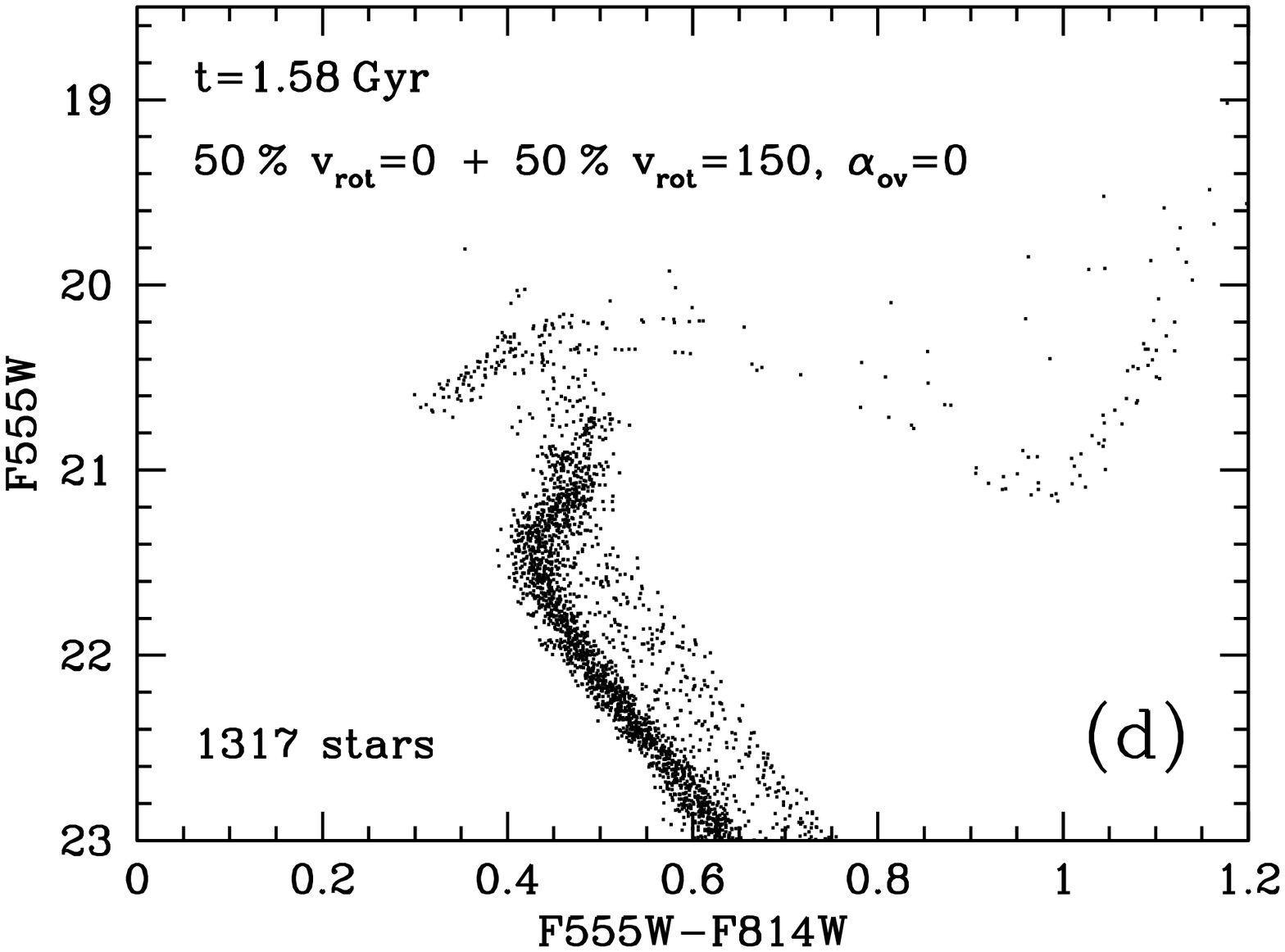}}
\resizebox{0.33\hsize}{!}{\includegraphics{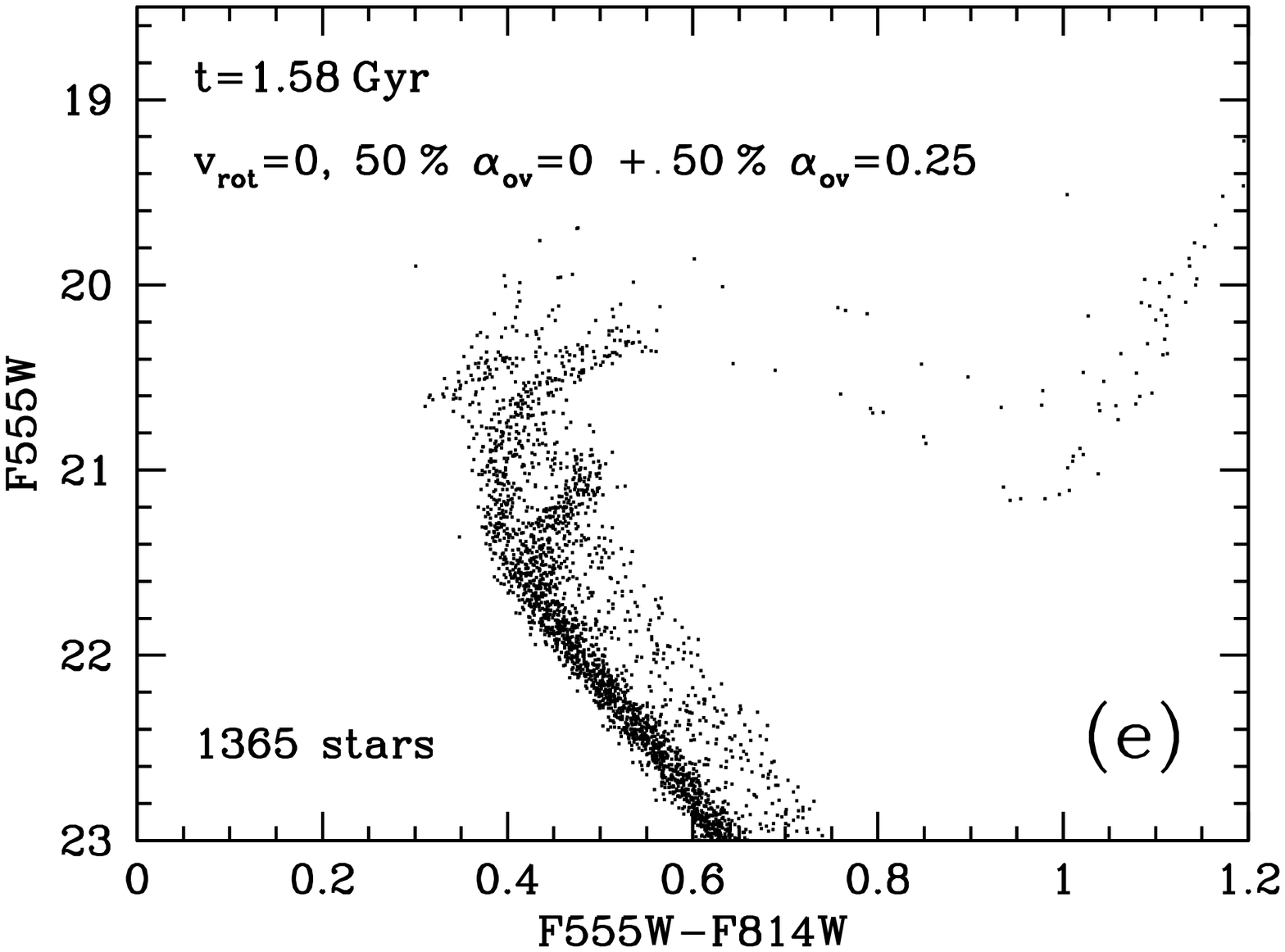}}
\resizebox{0.33\hsize}{!}{\includegraphics{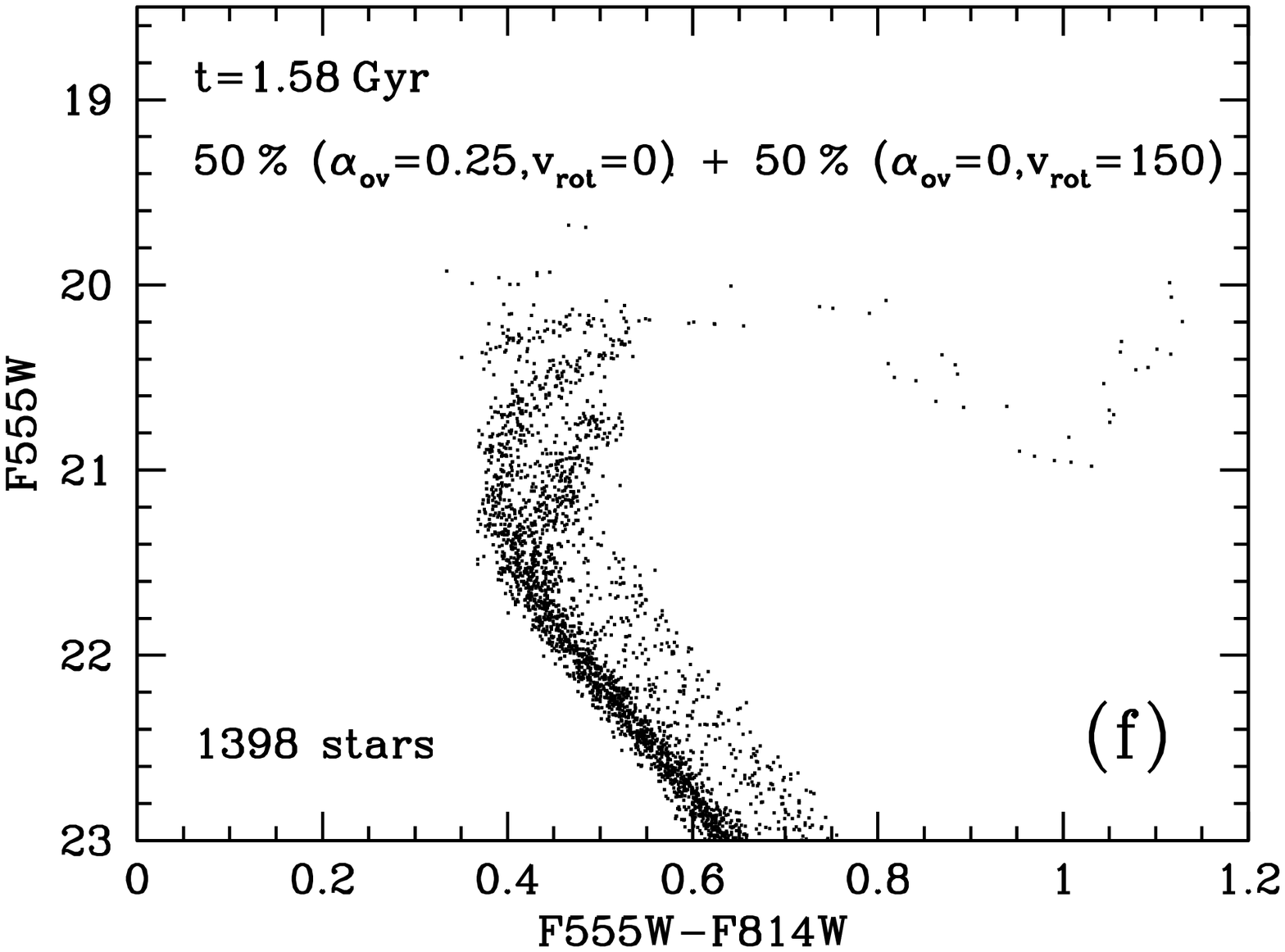}}
\resizebox{0.33\hsize}{!}{\includegraphics{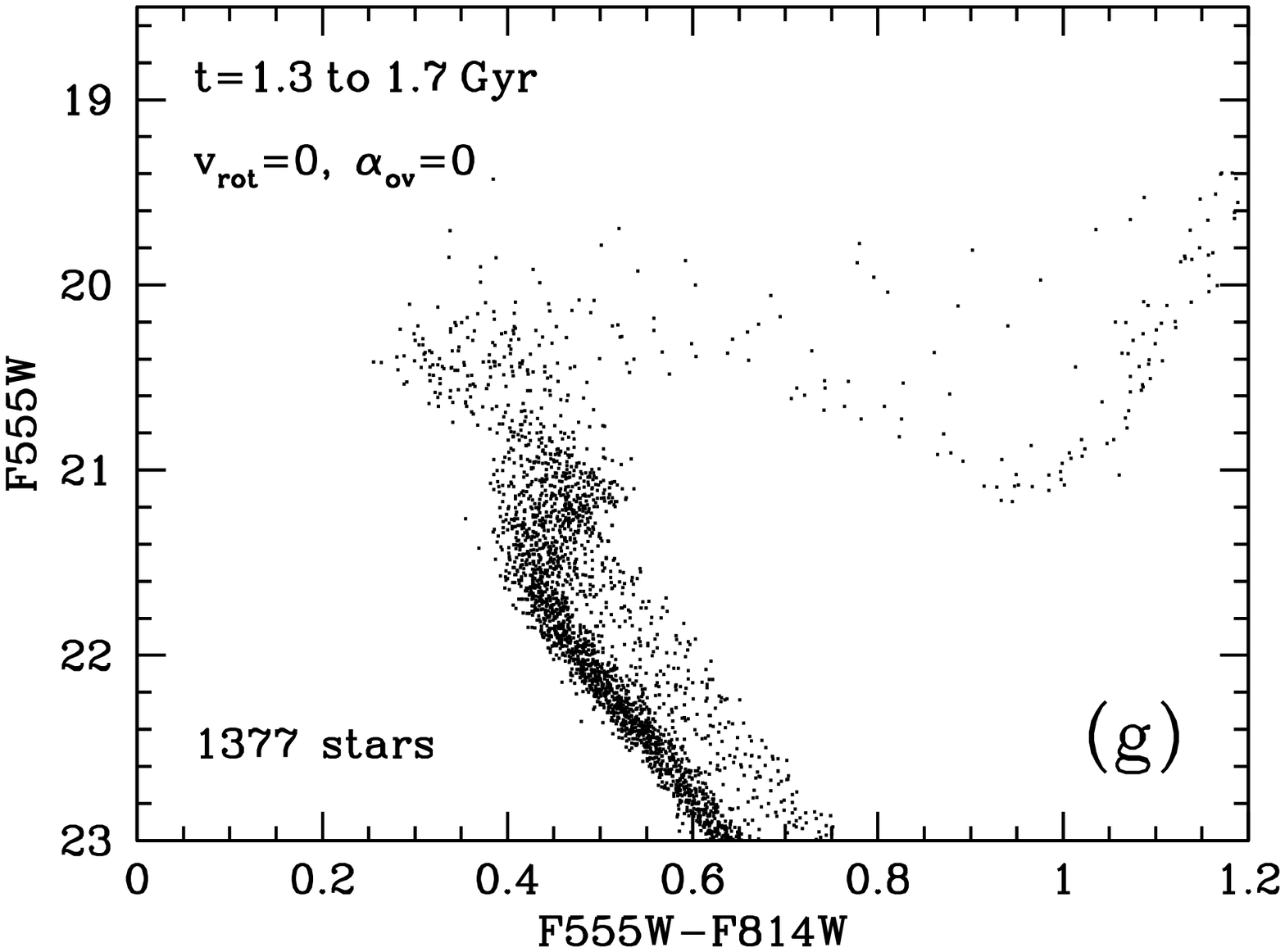}}
\resizebox{0.33\hsize}{!}{\includegraphics{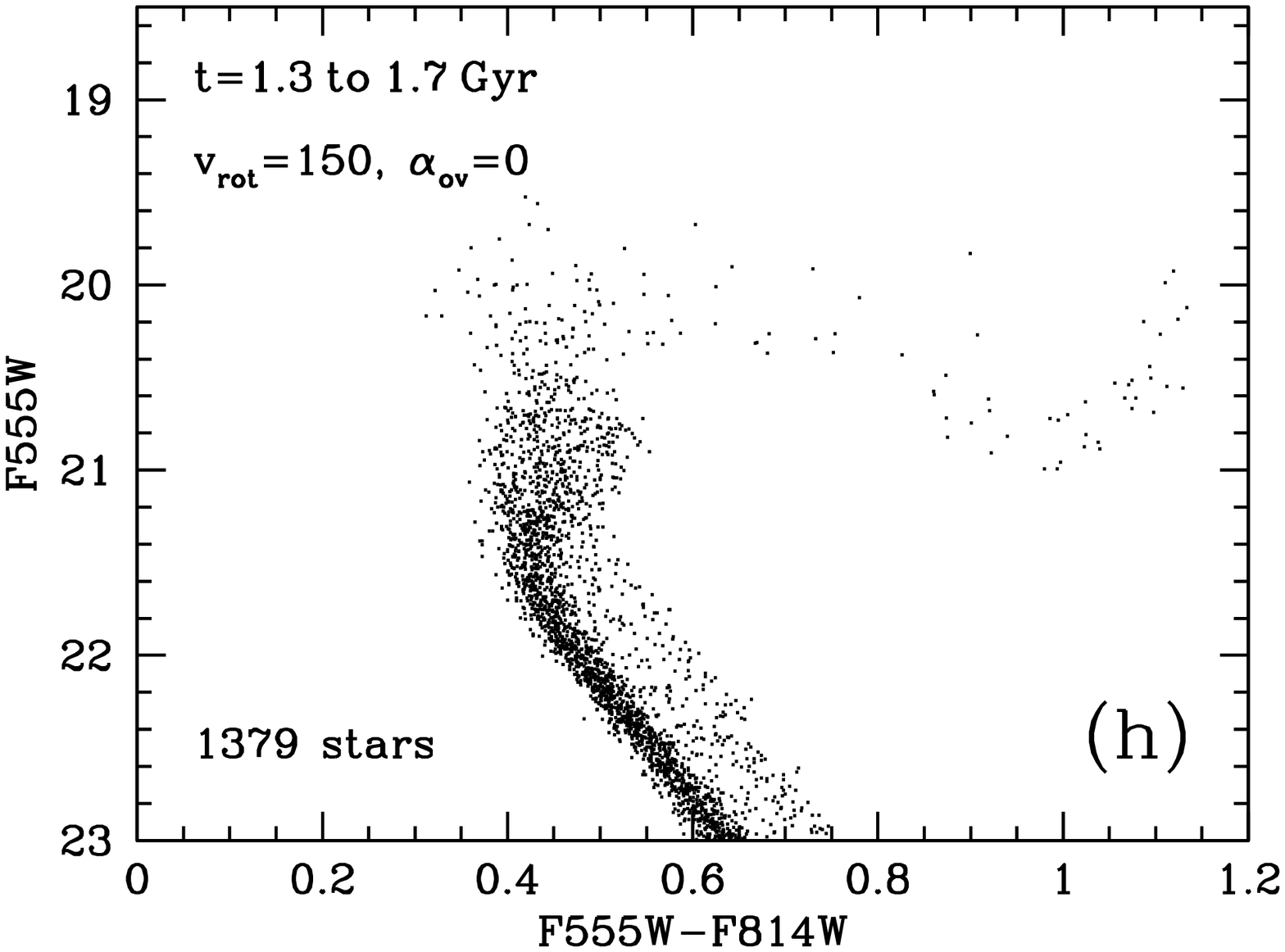}}
\resizebox{0.33\hsize}{!}{\includegraphics{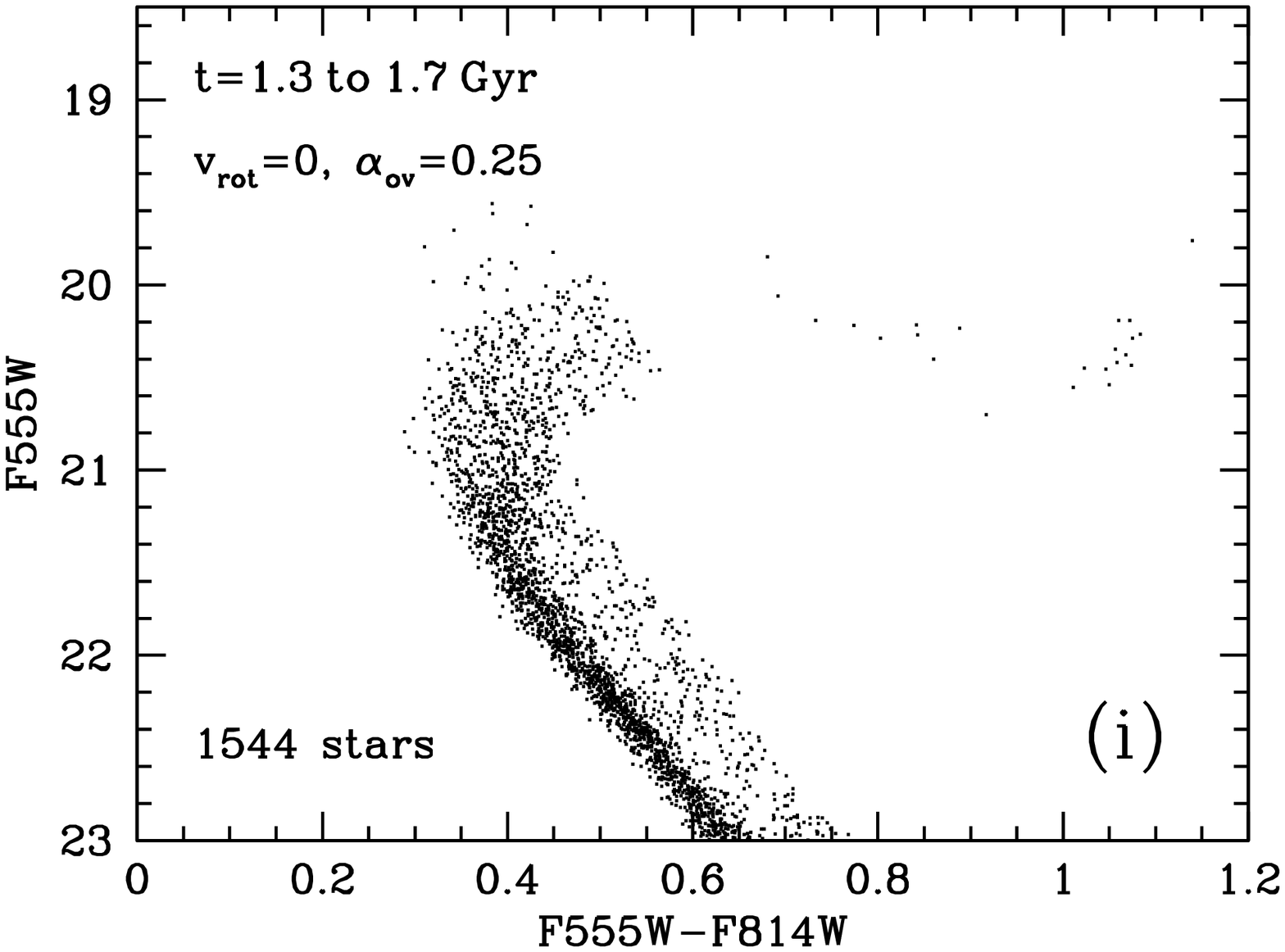}}
\resizebox{0.33\hsize}{!}{\includegraphics{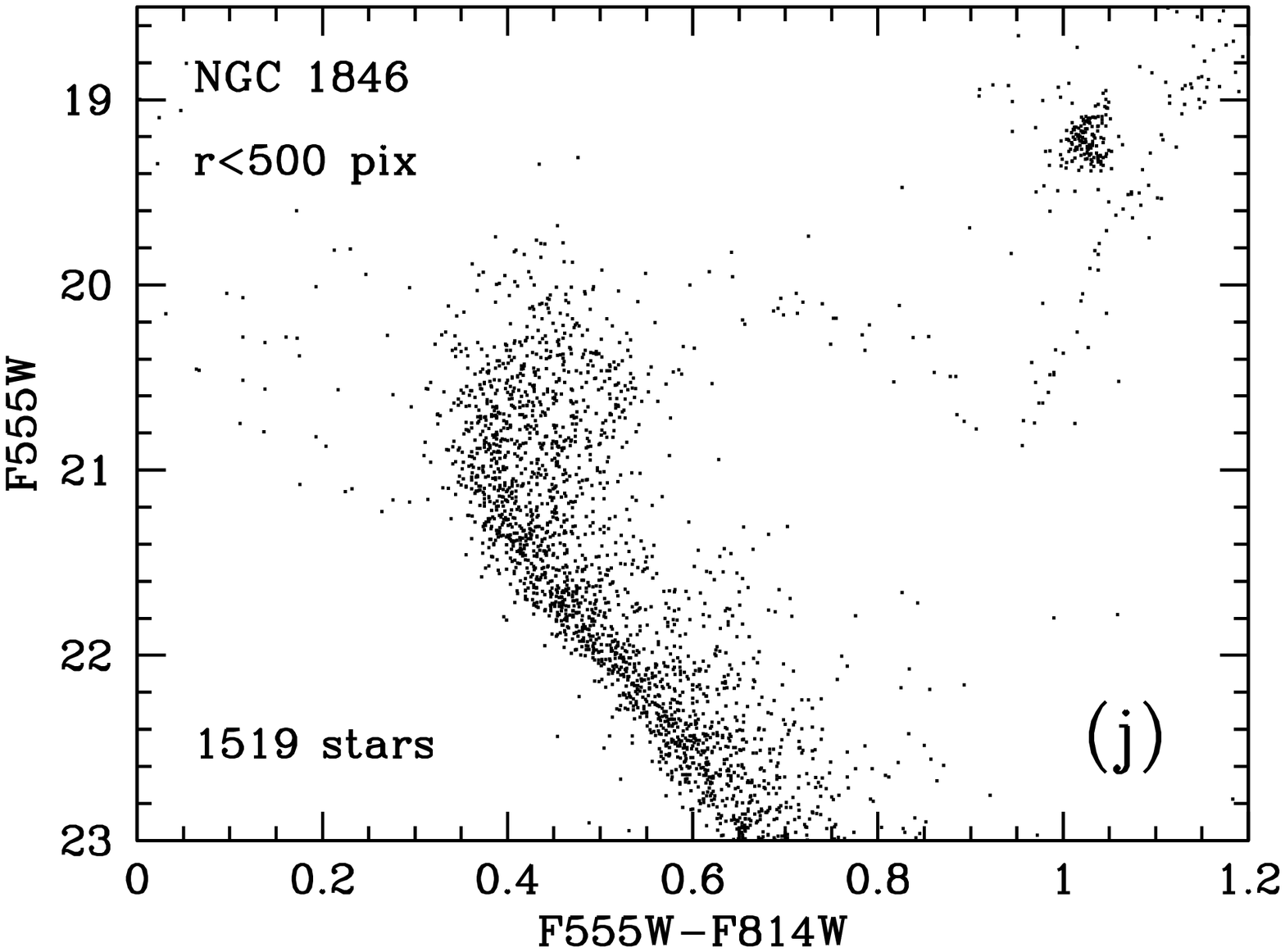}}
\caption{Panels (a)--(i) present synthetic CMDs illustrating the 
effect of different prescriptions regarding the distributions of
initial rotational velocities $v_{\rm rot}$, overshooting efficiency
$\alpha_{\rm ov}$, and ages $t$ among the clusters stars. All these
models assume a modest fraction of binaries and small photometric
errors. For comparison, panel (j) shows the ACS/WFC data for the
central region of the LMC star cluster NGC~1846
\citep[from][]{Goudfrooij_etal09}, with its well defined MMSTO. 
For each panel, we provide the number of objects in the turn-off
region defined by $19<{\rm F814W}<21.5$ and $0.2<{\rm
F555-F814W}<0.8$. The top row shows models for a single age and for
(a) no rotation without overshooting, (b) rotation without
overshooting, and (c) no rotation without overshooting. The second row
show mixed models, again for a single age: a mixture of (d) rotating
and non rotating stars, without overshooting, (e) stars with and
without overshooting, without rotation, and (f) non-rotating stars
with overshooting, together with rotating stars without
overshooting. It is evident that none of these cases produces the
peculiar ``golf club'' shape of the observed MMSTO. In the third row,
from (g) to (i), we have the same models as un the top row, but now
with the star formation spanning and interval of 400~Myr. The
similarity with the observed MMSTO is evident, especially in the panel
(i), corresponding to models with a moderate efficiency of convective
overshooting but no rotation.}
\label{fig_cmds}
\end{figure*}

Then, using the TRILEGAL code \citep{Girardi_etal05}, we simulate
populations in the CMD in which MMSTOs are usually observed, namely
the F555W vs. F555W$\,-\,$F814W (approximately $V$ vs. $\vi$) diagram
for the filters in the the ACS/WFC camera onboard HST. Such
simulations are shown in Fig.~\ref{fig_cmds}. The bolometric
correction tables in use are described in \citet{Girardi_etal08}. The
simulated photometry is displaced by the absolute distance modulus of
18.5~mag and a $V$-band foreground extinction 0.2~mag, which are
values typical of LMC star clusters. Then, photometric errors of
$\sigma=0.01$~mag are applied to the photometry, so as to help in the
visualisation of the densest parts of the CMDs without however
modifying their morphology in any appreciable way. All simulations
include a fraction of $20$~\% of binaries with mass ratios uniformly
distributed between 0.4 and 1. The total mass is assumed to be of
$2.4\times10^4$~\Msun, following a \citet{Chabrier01} log-normal IMF.

The models in the top row of Fig.~\ref{fig_cmds}, from (a) to (c), are
for a single age of 1.58~Gyr, and using each one of the 3 sets of
tracks presented in this work. These top panels illustrate the modest
effect that binaries have in these CMDs, in a way similar to the
synthetic CMDs already presented by many authours
\citep[e.g.][]{Bertelli_etal03, Kerber_etal07, Mackey_etal08}. 

Panels in the second row of Fig.~\ref{fig_cmds} represent attempts to
produce MMSTOs by adding stellar models with different properties but
again for the same single age. Panel (d) shows the case of models with
and without rotation and otherwise identical physics. This corresponds
to the situation favoured by \citet{BastiandeMink09} as being at the
origin of MMSTOs. As can be noticed, our results produce a
dramatically narrow MSTO, clearly very different from both the
observed one and those illustrated in \citet{BastiandeMink09}. 

Panel (e) instead explores the case in which there is a mixture of
stars with different efficiencies of core overshooting, but observed
at the same age. In this case the MSTO opens into a composite feature,
with two distinct MSTOs. However, the brighter MSTO (corresponding to
the overshooting models) is clearly much more extended to larger radii
(towards the top-right corner of the figure) than the fainter
MSTO. This again contrasts with observations, in which the MMSTOs are
seen to have about the same extension in the direction perpendicular
to the lower main sequence. Finally, the panel (f) shows the
combination of non-rotating stars with overshooting, together with
rotating stars without overshooting, so as to simulate an hypothetical
reduction of the extent of the overshooting region driven by
rotation. This combination produces CMDs slightly more similar to the
observed one, but does not reproduce its detailed shape; in
particular, it can be noticed that the fainter turn-off, corresponding
to rotating models without overshooting, is again less extended in the
direction perpendicular to the main sequence than the brighter
one. This discrepancy cannot be remediated imposing that the fast
rotators have some residual amount of overshooting, since this would
increase their lifetimes and move the fainter turn-off towards the
brighter one, creating a situation similar to the one in panel
(d). The adoption of faster rotational velocities in models without
overshooting would also move the faintest turn-off upwards in a
similar way.

The third row of Fig.~\ref{fig_cmds}, from (g) to (i), presents the
models derived from each set of homogeneous tracks, but now assuming
that star formation proceeded at a constant rate in the age interval
from 1.3 to 1.7~Myr. The similarity with the observed MMSTO is now
evident, especially in the panel (i), corresponding to models with a
single efficiency of overshooting but no rotation.

\section{Final considerations}
\label{sec_conclu}

The present work demonstrates that stellar rotation, as predicted by
detailed evolutionary models, produces features in CMDs which are
quite different from those illustrated in \citet{BastiandeMink09}. We
find that, despite the changes that rotation causes in the HR diagram
of evolutionary tracks, their longer lifetimes make the isochrones
derived from these tracks to have a MSTO position almost
indistinguishable from those derived from non-rotating models. It
follows that coeval isochrones derived from rotating plus non-rotating
tracks -- with otherwise identical physics -- are not able to describe
the detailed shape of the observed MMSTOs in Magellanic Cloud
clusters.

On the other hand, Fig.~\ref{fig_cmds} and the careful
CMD-reconstruction work by \citet{Rubele_etal10} and
\citet{Rubele_etal11} demonstrate that the other possible explanation
for the origin of MMSTOs -- namely prolonged histories of star
formation taking place in the clusters -- does provide excellent
qualitative and quantitative fits of the CMD in the clusters NGC~419
and NGC~1751. This despite these latter clusters present dual red
clumps in addition to MMSTOs, which represents a big additional
challenge to CMD-fitting algorithms. It is hardly conceivable, after
the results from this work, that a coeval population of stars with a
dispersion in their rotational velocities, could produce a CMD-fitting
that even approaches the quality reached in those CMD-fitting
works. {\em We are left with no better alternative, for the moment,
than accepting that prolonged star formation has probably occurred in
these clusters, although we are aware of the great challenges it
represents for dynamical models of stellar and cluster formation.}

We are also aware that stellar rotation -- and fast rotators -- occurs
{\em anyway} among the cluster stars. What is evinced by our models is
simply that the effect of fast rotation on CMDs is modest and more
likely similar to the colour and magnitude dispersion caused by
binaries, rather than the dramatic drawing of MMSTOs advocated by
\citet{BastiandeMink09}. Following our reasoning, the detection of
fast rotators at the MMSTOs is just expected, {\em and would not
support \citealt{BastiandeMink09}' hypothesis unless the fast rotators
are all observed at the reddest MSTO}. This is an aspect to be
tackled by future observations of rotation in cluster stars, as those
recently announced by \citet{Platais_etal10} for the open cluster Tr~20.

In conclusion, the explanation based on prolonged periods of SFH seems
to be the only surviving one, in this moment, to explain the origin of
MMSTOs in Magellanic Cloud clusters. Despite the perplexities and
difficulties it may cause to star formation theories, the presence of
dual red clumps in some clusters \citep{Girardi_etal09} and the high
quality of the results obtained from the modelling of their CMDs
\citep{Rubele_etal10, Rubele_etal11} strongly favour this
interpretation. Detection of chemical self-enrichment in clusters with
MMSTOs would likely be the decisive data to confirm or disprove this
explanation.

\section*{Acknowledgments}
We thank P. Goudfrooij for providing the NGC~1846 data plotted in
Fig.~\ref{fig_cmds}. The same data can be obtained from the
Multimission Archive at the Space Telescope Science Institute
(MAST). STScI is operated by the Association of Universities for
Research in Astronomy, Inc., under NASA contract NAS5-26555.  We thank
S. de Mink, N. Bastian and the anonymous referee for the constructive
comments, and A. Bressan for the help in building the
isochrones. L.G. acknowledges support from contract ASI-INAF
I/009/10/0.
%
%

\input{submitted.bbl}
%
\label{lastpage}
\end{document}

%% file: journalDefs.tex
\def\aj{AJ}%
\def\apj{ApJ}%
\def\apjl{ApJ}%
\def\apss{Ap\&SS}%
\def\aap{A\&A}%
\def\aaps{A\&AS}%
\def\mnras{MNRAS}%
\def\pasp{PASP}%